\documentclass[conference]{IEEEtran}
\IEEEoverridecommandlockouts

\usepackage[style=ieee,sortcites=true,sorting=none,maxbibnames=6, backend=biber,mincitenames=1,maxcitenames=2, dashed=false,]{biblatex}
\ifCLASSINFOpdf
\usepackage[pdftex]{graphicx}
\usepackage{tikz}
\usetikzlibrary{quantikz}
\usepackage{xcolor}
\usepackage{multirow}
\else
\fi

\usepackage{amsmath}
\usepackage{amsfonts} 
\usepackage{siunitx}
\usepackage{algorithmic} 

\usepackage{array}
\usepackage{url}
\usepackage{hyperref}
\usepackage{cleveref}
\usepackage{balance}

\begin{document}
\title{Solving Combinatorial Optimization Problems \\ with a Block Encoding Quantum Optimizer \thanks{This project is supported by the Federal Ministry for Economic Affairs and Climate Action on the basis of a decision by the German Bundestag through the project \textit{QuaST -- Quantum-enabling Services and Tools for Industrial Application}. QuaST aims to facilitate access to quantum-based solutions for optimization problems and to bridge the gap between business and technology.}}

\author{
    \IEEEauthorblockN{Adelina Bärligea, 
    Benedikt Poggel, 
    Jeanette Miriam Lorenz}
    \IEEEauthorblockA{Fraunhofer Institute for Cognitive Systems IKS, Munich, Germany
    \\\{adelina.baerligea, benedikt.poggel, jeanette.miriam.lorenz\}@iks.fraunhofer.de}
}

%\author{\IEEEauthorblockN{Adelina Bärligea}
%\IEEEauthorblockA{Fraunhofer Institute for\\ Cognitive Systems IKS\\ 80686 Munich, Germany\\ adelina.baerligea@iks.fraunhofer.de}
%\and
%\IEEEauthorblockN{Benedikt Poggel}
%\IEEEauthorblockA{Fraunhofer Institute for\\ Cognitive Systems IKS\\ 80686 Munich, Germany\\ benedikt.poggel@iks.fraunhofer.de}
%\and
%\IEEEauthorblockN{Jeanette Miriam Lorenz}
%\IEEEauthorblockA{Fraunhofer Institute for\\ Cognitive Systems IKS\\ 80686 Munich, Germany\\ jeanette.miriam.lorenz@iks.fraunhofer.de}}


\maketitle
\pagestyle{plain}

\begin{abstract}
% Background
In the pursuit of achieving near-term quantum advantage for combinatorial optimization problems, the Quantum Approximate Optimization Algorithm (QAOA) and the Variational Quantum Eigensolver (VQE) are the primary methods of interest, but their practical effectiveness remains uncertain. Therefore, there is a persistent need to develop and evaluate alternative variational quantum algorithms.
%Methods
This study presents an investigation of the Block ENcoding Quantum Optimizer (BENQO), a hybrid quantum solver that uses block encoding to represent the cost function. BENQO is designed to be universally applicable across discrete optimization problems. Beyond Maximum Cut, we evaluate BENQO's performance in the context of the Traveling Salesperson Problem, which is of greater practical relevance.
%Results
Our findings confirm that BENQO performs significantly better than QAOA and competes with VQE across a variety of performance metrics.
%Conclusions
We conclude that BENQO is a promising novel hybrid quantum-classical algorithm that should be further investigated and optimized to realize its full potential.

\end{abstract}

\IEEEpeerreviewmaketitle

\section{Introduction}
% BACKGROUND
The key technological challenge in the Noisy Intermediate-Scale Quantum (NISQ)~\cite{Preskill.2018} era is how to best make use of the currently available devices while accounting for their limited number of qubits, circuit depth, qubit connectivity, and the inherent noise. This requires not only identifying the type of problems to be solved, but also the corresponding algorithms that can run efficiently on NISQ hardware. 

The application field of combinatorial optimization~\cite{Korte.2018} attracts considerable attention from both the research community and industry, as any improvement to its state-of-the-art solutions would have a major economic and ecologic impact. While it is not widely believed that quantum computers will solve NP-complete problems in polynomial time, a possible advantage could arise in finding better or faster approximate solutions than current classical algorithms~\cite{Abbas.04.12.2023}. 

As of today, the prime candidates to bring this sort of advantage are Variational Quantum Algorithms (VQA)~\cite{Cerezo.2021}, which employ shallow parameterized quantum circuits to estimate the cost function while outsourcing the task of parameter optimization to classical computers. The most prominent examples are the Quantum Approximate Optimization Algorithm (QAOA)~\cite{Farhi.14.11.2014} and the Variational Quantum Eigensolver (VQE)~\cite{Peruzzo.2014}.
As research on advancing these two established solution methods and exploring their applications rapidly grows~\cite{Blekos.15.06.2023, Tilly.2022}, several limitations have also become apparent~\cite{Bittel.2021, Cerezo2023}, with the most challenges arising from hardware noise~\cite{StilckFrana2021, Wang2021, GonzalezGarcia.2022}. To overcome current constraints and improve quantum computational capabilities in the long run, novel quantum algorithms and heuristics should be explored, even without theoretical performance guarantees~\cite{Abbas.04.12.2023}.

% RESEARCH OBJECTIVES
Therefore, we investigate and evaluate a new hybrid quantum optimization algorithm introduced by Meli et al.~\cite{Meli.11.06.2023}, which has demonstrated promising results on weighted Maximum Cut (MaxCut) and Ising problems, significantly outperforming QAOA. It employs a new solution paradigm that uses block encoding to represent the Hermitian cost operator of an optimization problem within a quantum circuit, combined with the principle of implicit measurement~\cite{Nielsen.2016} to efficiently obtain cost function values. To expand its application potential and provide more comprehensive performance benchmarks, we embed it into a full framework called BENQO (Block ENcoding Quantum Optimizer). Enhancing the original concept, BENQO can universally be applied to all quadratic unconstrained binary optimization (QUBO) problems.

% METHODOLOGY & RESULTS
In this study, we turn our attention to the Traveling Salesperson Problem (TSP)~\cite{Korte.2018} -- a task with a more immediate practical relevance -- and use it as the primary test case for the newly established, streamlined framework. Building on previous results for the MaxCut problem~\cite{Meli.11.06.2023}, our work presents new, promising results for the TSP, which demonstrate the viability of BENQO. As we focus on the paradigm of (universal) gate-based quantum computing, we conduct a comparative analysis of BENQO against the two most prominent VQAs, QAOA and VQE. Besides examining aspects such as computational complexity and solution quality, we complement the experiments with a first qualitative analysis of the algorithms' loss landscapes and runtime characteristics. Ultimately, our study provides valuable insights into BENQO's performance under practical constraints. 

% SUMMARY OF STRUCTURE
The remainder of this paper is structured as follows: \Cref{sec:background} summarizes basic notions of Ising and QUBO problems, VQAs, and the TSP. \Cref{sec:BENQO} details the algorithm and its parameter optimization routine and gives a theoretical complexity analysis. The experimental results, including loss landscapes and the analysis of solution quality and runtime for weighted Maximum Cut and the TSP, are presented in \Cref{sec:results}. Finally, \Cref{sec:discussion} summarizes and discusses our findings, and \Cref{sec:outlook} outlines future research directions.

\section{Background and Related Work}
\label{sec:background}
\subsection{Ising and QUBO Problems}
A system of $n\in\mathbb{N}$ spin particles can be described by the $2^n\times2^n$ Ising Hamiltonian  
\begin{equation}
    \mathbf{C}=\sum_{i=1}^n \mathcal{C}_{i i} \mathbf{Z}_i+\sum_{i<j} \mathcal{C}_{i j} \mathbf{Z}_i \mathbf{Z}_j.
\label{eq:IsingCostFunction}
\end{equation}
Here, $\mathbf{Z}_i$ denotes the Pauli-$\mathbf{Z}$ operator with eigenvalues $\pm1$ acting on the $i^{\mathrm{th}}$ particle, and $\mathbf{C}_{ij}$ corresponds to the interaction energy between particles $i$ and $j$. For a system in state $|\boldsymbol{q}\rangle = \bigotimes_{i=1}^n|q_i\rangle$ with $q_i\in\{0,1\}$, the total energy $\langle \mathbf{C}\rangle := \langle\boldsymbol{q}| \mathbf{C}|\boldsymbol{q}\rangle$ is defined as
\begin{equation}
\langle \mathbf{C}\rangle=\sum_{i=1}^n(-1)^{q_i}\: \mathcal{C}_{i i}+\sum_{i<j}(-1)^{q_i+q_j}\: \mathcal{C}_{i j}.
\label{eq:IsingCostExpectation}
\end{equation}
The resulting optimization problem of finding the ground state $|\boldsymbol{q}^*\rangle$ is equivalent to minimizing the expectation \eqref{eq:IsingCostExpectation}, which is known as the Ising problem. Note that, when setting all self-connections $\mathcal{C}_{i i}=0$, this corresponds to the weighted MaxCut problem. It has been demonstrated that many NP-hard discrete optimization problems of practical relevance can be described within this formalism~\cite{Lucas.2014}.

A mathematically equivalent formulation to Ising problems are quadratic unconstrained binary optimization (QUBO) problems
\begin{equation}
    \min_{\mathbf{x}} \mathbf{x}^{\top} \mathbf{Q}\: \mathbf{x}.
\label{eq:QUBOproblem}
\end{equation}
with binary decision variables $\mathbf{x}\in\{0,1\}^n$~\cite{Fu1986}. In order to translate a QUBO problem \eqref{eq:QUBOproblem} -- defined fully by the matrix $\mathbf{Q}$ -- into its corresponding Ising form \eqref{eq:IsingCostFunction} defined by the Ising weights $\mathcal{C}_{i j}$, one needs to transform the binary variables $x$ into spin variables $z\in \{\pm1\}$ via $x=(z+1)/2$ such that $0$ is mapped to $+1$ and $1$ to $-1$, and then replace the spin variables with Pauli-$\mathbf{Z}$ operators. The resulting Ising Hamiltonian in terms of a symmetric QUBO matrix ${\mathbf{Q}=\{Q_{ij}\}_{i,j=1,\dots,n}}$ then reads
\begin{equation}
\mathbf{C} =\sum_{i<j} \frac{Q_{i j}}{2} Z_{i} Z_{j}+\sum_{i=1}^{n}\left(\sum_{j=1}^{n} \frac{Q_{i j}}{2}\right) Z_{i}+\text { const. },
\label{eq:QUBO2Ising}
\end{equation}
for which the constant offset can be ignored when computing the cost. An exemplary mapping between the QUBO and MaxCut formulation can be found in Ref.~\cite{Barahona.1989}.

\subsection{Variational Quantum Algorithms}
VQAs~\cite{McClean.2016, Cerezo.2021} can find solutions to combinatorial optimization problems by constructing a parameterized quantum circuit that prepares the candidate state $|\Psi(\boldsymbol{\theta})\rangle$ and then solves the minimization problem
\begin{equation}
\min_{\boldsymbol{\theta}}\ \mathcal{L}(\boldsymbol{\theta}) \quad\mathrm{with}\quad \mathcal{L}(\boldsymbol{\theta}):=\langle\Psi(\boldsymbol{\theta})|\mathbf{C}| \Psi(\boldsymbol{\theta})\rangle.
\label{eq:parametrizedCostFunction}
\end{equation}
With optimal parameters $\boldsymbol{\theta}^*$, $|\Psi(\boldsymbol{\theta^*})\rangle$ is a good approximation to the actual ground state $|\boldsymbol{q}^*\rangle$.

Currently, the two most exhaustively studied variational quantum algorithms to run on NISQ devices are the QAOA and the VQE. Both are generally hoped to bring some sort of quantum advantage in the near term. As BENQO builds upon and directly competes with these algorithms, we give a concise review here:

%\subsubsection{Variational Quantum Eigensolver}
The VQE, first proposed in 2014~\cite{Peruzzo.2014, Tilly.2022}, is grounded on the variational principle
\begin{equation}
    E_0 \leq \frac{\langle\psi|\mathbf{H}| \psi\rangle}{\langle\psi | \psi\rangle},
\label{eq:variationalprinciple}
\end{equation}
that the ground state energy $E_0$ associated with a given Hamiltonian $\mathbf{H}$ is upper bounded by its expectation value with regard to a trial state $| \psi\rangle$. The objective of a VQE is then to find a parametrized ansatz $\mathbf{U}(\boldsymbol{\theta})$, such that $| \psi\rangle=\mathbf{U}(\boldsymbol{\theta})|\mathbf{0}\rangle$ best approximates the eigenvector of $\mathbf{H}$ corresponding to the smallest eigenvalue $E_0$, namely
\begin{equation}
    E_{\mathrm{VQE}}=\min _{\boldsymbol{\theta}}\ \langle\mathbf{0}|\:\mathbf{U}^{\dagger}(\boldsymbol{\theta})\ \mathbf{H}\ \mathbf{U}(\boldsymbol{\theta})\: | \mathbf{0}\rangle.
\label{eq:VQEobjective}
\end{equation}
In principle, any ansatz $\mathbf{U}(\boldsymbol{\theta})$ can be chosen between problem-inspired and hardware-efficient ones~\cite{Cerezo.2021}. While research on the specific application of VQE on the MaxCut problem exists only sporadically~\cite{Moll.2018}, there are some studies showing its comparatively good performance when applied to heavily constrained optimization problems such as the TSP~\cite{Khumalo.2021, Palackal.19.04.2023}.

%\subsubsection{Quantum Approximate Optimization Algorithm}
Another popular approach is the QAOA~\cite{Farhi.14.11.2014}, for which there is currently a drastic expansion in research activity~\cite{Blekos.15.06.2023}. The QAOA can be seen as a trotterized version of adiabatic quantum optimization~\cite{farhi2000}. It uses a phase shift operator $\mathbf{U}_P(\gamma)=e^{-i \gamma \mathbf{H}_P}$ with the problem Hamiltonian $\mathbf{H}_P$, and a mixing operator $\mathbf{U}_M(\beta)=e^{-i \beta \mathbf{H}_M}$, where the default choice is a sum of Pauli-$\mathbf{X}$ matrices on each qubit $\mathbf{H}_M=\sum_{i=1}^n \mathbf{X}_i$. The resulting unitary combines these operators in $p$ layers, parameterized by the $2p$ angles $\boldsymbol{\gamma}=\{\gamma_i\}_{i=1}^p$ and $\boldsymbol{\beta}=\{\beta_i\}_{i=1}^p$: 
\begin{equation}
\mathbf{U}(\boldsymbol{\gamma}, \boldsymbol{\beta}) =
\mathbf{U}_M\left(\beta_p\right) \mathbf{U}_P\left(\gamma_p\right) \dots \mathbf{U}_M\left(\beta_1\right) \mathbf{U}_P\left(\gamma_1\right),
\end{equation}
The unitary is then applied to an initial state $|\psi_0\rangle$, typically the uniform superposition. The resulting objective can be calculated the same way as in \cref{eq:VQEobjective}.

To date, the MaxCut problem is the primary use case examined with QAOA, establishing it as a standard quantum benchmark for this problem~\cite{Farhi.14.11.2014,Blekos.15.06.2023}. However, when extending QAOA to tackle more complex problems, such as the TSP, it is notably less efficient than the VQE~\cite{Khumalo.2021, Palackal.19.04.2023}. While there have been promising efforts to improve this performance~\cite{Ruan.2020, Xie.17.08.2023}, our study focuses on the standard version outlined above, which aligns with the TSP problem formulation described below.

\subsection{Traveling Salesperson Problem}
\label{ssec:TravelingSalesmanProblem}
The TSP, a renowned NP-hard problem in combinatorial optimization~\cite{Karp1972}, has significant applications from logistics to route planning. Despite its extensive study, classical solutions have not yet met the problem's theoretical inapproximability bounds~\cite{Ausiello1999}, which is why it is especially suitable to be tested on quantum solvers~\cite{Abbas.04.12.2023}, such as hybrid VQAs.

The TSP is typically defined on a fully-connected, undirected graph $\mathcal{G}=(\mathcal{V}, \mathcal{E})$ with $\mathcal{V}$ the set of $n$ nodes and $\mathcal{E}$ a set of weighted edges with a "distance" $\omega_{ij}$ assigned to each edge (connecting the $i^{\text{th}}$ and $j^{\text{th}}$ node). The goal is to find a Hamiltonian cycle with a minimal total distance that passes through each node of the graph. 

In this study, the formulation proposed by Lucas~\cite{Lucas.2014} is considered, with a QUBO cost function
\begin{multline}
    C = \sum_{i, j=1}^n \omega_{i j} \sum_{\alpha=1}^n x_i^\alpha x_j^{\alpha+1} \\ +P \sum_{i=1}^n\left(1-\sum_{\alpha=1}^n x_i^\alpha\right)^2+P \sum_{\alpha=1}^n\left(1-\sum_{i=1}^n x_i^\alpha\right)^2.
\label{eq:TSPformulation}
\end{multline}
The paths are described by $n^2$ binary decision variables $x_i^{\alpha}$, marking whether node $i$ is visited at time step $\alpha$. This formulation corresponds to a one-hot encoding of the equivalent integer problem and directly translates into an Ising model \eqref{eq:IsingCostFunction}.
The two necessary permutational constraints
\begin{itemize}
    \item $\sum_i x_i^{\alpha} = 1\quad \forall\ \alpha=1,\dots,n$ \\ (at each step, only one node can be visited)
    \item $\sum_{\alpha} x_i^{\alpha} = 1\quad \forall\ i = 1,\dots ,n$ \\ (each node should appear only once in the cycle)
\end{itemize}
are included as quadratic penalty terms~\cite{Glover.2022} with a penalty factor $P$. The total length of the chosen cycle is computed in the first term of \cref{eq:TSPformulation}. In order to reach feasible solutions, $P$ needs to be larger than a minimum penalty $P_{\min}$ which is known to be lower bounded by $\max(\omega_{ij})$~\cite{Lucas.2014}. To choose $P$ appropriately, \textsc{Qiskit}'s~\cite{Qiskit} default strategy is adopted, using the absolute value range of the given unconstrained quadratic cost expression \eqref{eq:QUBOproblem}. Considering that the product of any two binary variables $x_i^{\alpha}$ in the first term of \cref{eq:TSPformulation} still has bounds $0$ and $1$, this penalty factor can be calculated as $P=n\cdot\sum_{i, j} \omega_{i j}$. In fact, our experiments revealed that values much smaller than this, as suggested in~\cite{Palackal.19.04.2023}, lead to significantly fewer feasible solutions.

Lastly, note that the TSP can be reduced to ${(n-1)^2}$ variables by fixing the starting point of each cycle, setting ${x_1^1=1}$, ${x_1^{\alpha}=0\ \forall\alpha\neq1}$, and ${x_i^{i}=0\ \forall i\neq1}$, effectively eliminating the cyclic permutation symmetry of the problem.

\section{The BENQO Algorithm}
\label{sec:BENQO}
This section is based primarily on the algorithm UQMaxCutAndIsing introduced and described in more detail in~\cite{Meli.11.06.2023}. To evaluate the algorithm's applicability to combinatorial optimization problems beyond MaxCut and Ising, such as the TSP, we formulate it in a universal framework named BENQO (Block ENcoding Quantum Optimizer). It is computationally streamlined and independent of the specific problem instance, and can therefore be applied to any QUBO problem.

\subsection{Algorithm Outline}
\label{ssec:Algorithm Outline}
To solve the optimization problem stated in \cref{eq:parametrizedCostFunction} using a quantum computer, Meli et al.~\cite{Meli.11.06.2023} propose to embed the Hermitian cost matrix $\mathbf{C}$ defined in \eqref{eq:IsingCostFunction} into a larger (${2^{n+1}\times2^{n+1}}$) unitary
\begin{equation}
    \mathbf{U}:=\left[\begin{array}{rr}
    \sin (\hat{\mathbf{C}}) & \cos (\hat{\mathbf{C}}) \\
    \cos (\hat{\mathbf{C}}) & -\sin (\hat{\mathbf{C}})
    \end{array}\right] \quad \mathrm{with}\quad  \hat{\mathbf{C}}:=\mathbf{C} / K.
\label{eq:blockencoding}
\end{equation}
This encoding requires one additional qubit. $K$ is chosen such that the entries of $\mathbf{C}$ lie within $[-\frac{\pi}{2},+\frac{\pi}{2}]$. Then, $\mathbf{U}$ is a monotone, bijective mapping of the cost $\mathbf{C}$. This technique is known as block encoding~\cite{Low2019}, as one can retrieve the upper left block of \eqref{eq:blockencoding} by setting the additional qubit to state $|0\rangle$, such that 
\begin{equation}
    \langle\mathbf{U}\rangle:=\langle 0, q|\mathbf{U}| 0, q\rangle \equiv \sin \langle\hat{\mathbf{C}}\rangle.
\label{eq:UtoC}
\end{equation}

Applying a few transformations to the unitary matrix \eqref{eq:blockencoding} using trigonometric and rotational identities, one arrives at
\begin{equation}
\begin{aligned}
& \mathbf{U}(\mathbf{C}, K) =
%\mathbf{R}_{\mathrm{y}}(-2\langle\widehat{\mathbf{C}}\rangle) \cdot \mathbf{X} \otimes \mathbf{I}^{\otimes \mathrm{n}} = 
\\
& \quad \prod_{i=1}^n\quad\: \mathbf{X}^{q_i} \cdot \mathbf{R}_{\mathrm{y}}\left(-2 \hat{\mathcal{C}}_{i i}\right) \cdot \mathbf{X}^{q_i} \cdot \\
& \prod_{1 \leq i<j \leq n} \mathbf{X}^{q_i+q_j} \cdot \mathbf{R}_{\mathrm{y}}\left(-2 \hat{\mathcal{C}}_{i j}\right) \cdot \mathbf{X}^{q_i+q_j} \cdot \mathbf{X} \otimes \mathbf{I}^{\otimes n}
\end{aligned}
\label{eq:unitaryoperator}
\end{equation}
with the expectation value \eqref{eq:IsingCostExpectation} inserted. As this corresponds to a product of simple unitary transformations, it can be implemented efficiently on a gate-based quantum computer. 

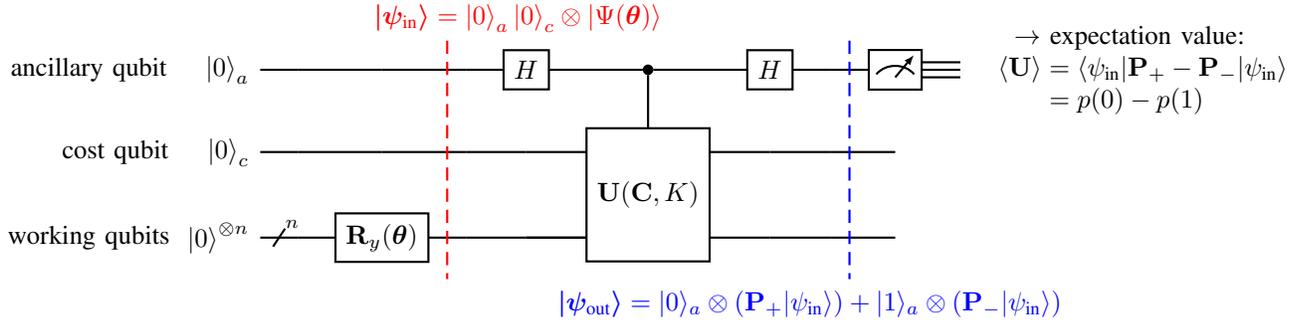
\begin{figure*}[!t]
    \centering
\begin{quantikz}
    & \text{ancillary qubit}\qquad & \lstick{$\ket{0}_a$} & \qw & \qw & \qw & \gate{H} & \ctrl{1} & \gate{H} & \qw & \meter{} & \rstick{$\rightarrow$ expectation value:\qquad \\ $\quad\langle\mathbf{U}\rangle = \langle\psi_{\text{in}}|\mathbf{P_+ - P_-}|\psi_{\text{in}}\rangle$\\ $\quad= p(0)-p(1)\quad\: $} \qwbundle[alternate]{} \\
    & \text{cost qubit} & \lstick{$\ket{0}_c$} & \qw & \qw & \qw & \qw & \gate[2]{\mathbf{U}(\mathbf{C}, K)} & \qw & \qw & \qw \\
    & \text{working qubits}\qquad & \lstick{$\ket{0}^{\otimes n}$} & \qwbundle{n} & \gate{\mathbf{R}_y(\boldsymbol{\theta})} \slice[label style={pos=-0.2, anchor=north, color=red}]{$\qquad \qquad \quad\ \boldsymbol{\ket{\psi_{\text{in}}}}=\ket{0}_a\ket{0}_c\otimes |\Psi(\boldsymbol{\theta})\rangle$} & \qw & \qw & \qw & \qw & \qw \slice[label style={pos=1., anchor=north, color=blue}, style={draw=blue}]{$\boldsymbol{|\psi_{\text{out}}\rangle} =|0\rangle_a \otimes\left(\mathbf{P_{+}}|\psi_{\text{in}}\rangle\right) +|1\rangle_a \otimes\left(\mathbf{P_{-}}|\psi_{\text{in}}\rangle\right)\qquad\quad $} & \qw
\end{quantikz}
    \caption{Circuit Diagram for the quantum optimizer described in Section \ref{ssec:Algorithm Outline} based on Ref.~\cite{Meli.11.06.2023}. The state $\boldsymbol{\ket{\psi_{\text{in}}}}$, parameterized by the angles $\boldsymbol{\theta}\in \mathbb{R}^n$ is prepared in a first step. Then, the Hadamard test is applied to the unitary operator $\mathbf{U}(\mathbf{C}, K)$ in order to project the resulting quantum system onto its eigenspace via $\mathbf{P_{\pm}}=(\mathbf{I_n\pm U})/2$. This allows to “measure the operator” in terms of the probabilities $p(0) = \lVert \mathbf{P_{+}}|\psi_{\text{in}}\rangle \rVert^2$ and $p(1) = \lVert \mathbf{P_{-}}|\psi_{\text{in}}\rangle \rVert^2$.}
    \label{fig:CircuitStructure4BENQO}
\end{figure*}

Finally, to obtain an estimate for the expectation $\langle\mathbf{U}\rangle$, Meli et al.~\cite{Meli.11.06.2023} propose to apply the principle of implicit measurement~\cite{Nielsen.2016} to conditionally project the eigenvalues of $\mathbf{U}$ onto the amplitudes of another additional ancillary qubit (see Figure \ref{fig:CircuitStructure4BENQO}). From its measurements,  $\langle\mathbf{U}\rangle\equiv p(0)-p(1)$ can be calculated, with $p(x)$ the probability of measuring $x=\{0,1\}$ respectively. The exact cost value can therefore directly be computed as $\langle\mathbf{C}\rangle=K\cdot \arcsin{\langle\mathbf{U}\rangle}$. With this technique, the number of samples needed to accurately estimate the cost value is independent of the number of working qubits in the ansatz, avoiding sampling the full state.

The overall circuit is displayed in Figure \ref{fig:CircuitStructure4BENQO}. The chosen ansatz
\begin{equation}
    |\Psi(\boldsymbol{\theta})\rangle=\bigotimes_{i=1}^n\mathbf{R}_{\mathrm{y}}(\theta_i)\ |0\rangle
\label{eq:rotationalansatz}
\end{equation}
with parameters $\boldsymbol{\theta}=\{\theta_i\}_{i=1,\dots n}\in \mathbb{R}^n$ is simple, consisting of one layer of single rotation gates applied to each working qubit in state $|0\rangle$. 

\subsection{Parameter Optimization}
\label{ssec:ParameterOptimization}
To optimize the parameters in the objective \eqref{eq:parametrizedCostFunction}, Meli et al.~\cite{Meli.11.06.2023} use a normalized gradient descent (NGD) method~\cite{Suzuki.} with an exponentially decaying step size, such that the update rule reads
\begin{equation}
\begin{aligned}
    & \boldsymbol{\theta^{(k)}}\ \rightarrow \\ & \boldsymbol{\theta^{(k-1)}} - \sqrt{\frac{\pi}{2}\:n}\cdot \exp \left(-\frac{4 k^2}{k_{\max }^2}\right) \cdot \frac{\nabla_{\boldsymbol{\theta}} \mathcal{L}\left(\boldsymbol{\theta}^{(k-1)}\right)}{\left\|\nabla_{\boldsymbol{\theta}} \mathcal{L}\left(\boldsymbol{\theta}^{(k-1)}\right)\right\|_2}
\label{eq:normalizedgradientdescent}
\end{aligned}
\end{equation}
with a chosen maximal number of steps $k_{\max}$. This technique of normalizing the gradient is supposed to help to escape local minima in the loss landscapes of VQAs~\cite{Suzuki.}, while the decreasing step size omits the need to determine any stopping criterion for the algorithm. A more detailed explanation for this choice can be found in~\cite{Meli.11.06.2023}.

To evaluate the gradient in \eqref{eq:normalizedgradientdescent}, we make use of the parameter shift rule~\cite{Mitarai2018}, defined as
\begin{equation}
\frac{\partial}{\partial \theta_i} \mathcal{L}(\boldsymbol{\theta})= \frac{\mathcal{L}(\theta_{i,+})-\mathcal{L}(\theta_{i,-})}{2} 
\label{eq:parametershiftrule}
\end{equation}
with $\theta_{i,\pm}:= (\theta_1,\dots, \theta_i\pm\frac{\pi}{2},\dots,\theta_n)$, which allows for an analytic evaluation of the gradient of a quantum circuit~\cite{Schuld.2019} via $2n$ function calls.

\subsection{Resource and Complexity Analysis}
\label{ssec:ressourceandcomplexity}
For an Ising-type problem with $n$ nodes and $m$ connections, the unitary operator $\mathbf{U}(\mathbf{C}, \mathrm{K})$ defined in \cref{eq:unitaryoperator} requires ${4m-2n}$ \textsc{cnot} gates and $n$ single qubit rotations. In Ref.~\cite{Meli.11.06.2023}, it is shown that in order to control $\mathbf{U}$ as is required for the circuit displayed in Figure \ref{fig:CircuitStructure4BENQO}, \cref{eq:unitaryoperator} can simply be expressed as a product of \textsc{cnot}s and controlled rotations, which in turn can generally be decomposed as 
\begin{equation}
\left[\mathbf{R}_y(\theta)\right]^a=\mathbf{X}^a \cdot R_y(-\theta / 2) \cdot \mathbf{X}^a \cdot \mathbf{R}_y(\theta / 2),
\end{equation}
where $a$ is the control value. This decomposition further doubles the number of single qubit rotations for each connection and increases the number of \textsc{cnot} gates by $2m$. 

Table \ref{tab:complexity} summarizes these findings and extends the analysis of~\cite{Meli.11.06.2023} to a broader comparison of BENQO with yet another state-of-the-art solver, namely VQE. We include the rotational gates of the ansatz into the resource analysis and present the circuit complexity in a more comparable way, depending only on $n$, using that fully-connected, undirected graphs have ${m=n(n+1)/2}$ edges (when counting self-connections). Especially for practically relevant optimization problems with hard constraints, the corresponding problem graphs are often relatively dense. Still, one can verify that BENQO is more gate-efficient than QAOA as soon as $p\geq3$.

For the VQE, a simple two-local ansatz (a layer of rotational gates followed by a linear entangling layer using controlled-$\mathbf{Z}$ gates) was chosen, such that it has the same number of parameters as BENQO. The controlled-$\mathbf{Z}$ gates of the ansatz can be decomposed using $\mathbf{Z}=\mathbf{H}\mathbf{X}\mathbf{H}$, resulting in the additionally necessary $2(n-1)$ Hadamard gates in Table \ref{tab:complexity}. As the full circuit of a VQE only consists of the chosen ansatz for preparing the solution states, the quantum resources required are much smaller than for QAOA and BENQO, whose circuits are problem-dependent and grow with the number of qubits. This advantage, however, is partly compensated by BENQO'S number of Pauli measurement bases needed to get an estimate for $\langle C\rangle$, which for VQE and QAOA is $m$ (also leading to a higher variance). In contrast, BENQO only need a single one.

\begin{table*}[!t]
%% increase table row spacing, adjust to taste
\renewcommand{\arraystretch}{1.25}
\caption{Circuit Resource and Complexity Comparison of BENQO with a conventional QAOA of $p$ layers and a VQE$^*$ ($^*$with a simple one-layered two-local ansatz circuit) for Ising cost functions defined on a fully-connected, undirected graph $\mathcal{G}=(\mathcal{V}, \mathcal{E})$, \\ with $n\equiv\|\mathcal{V}\|$ the number of nodes, and $m\equiv\|\mathcal{E}\|=n(n+1)/2$ the number of edges (including self-connections).}
\label{tab:complexity}
\centering
\begin{tabular}{|r||c|c|c|}
\hline
Algorithm & \textbf{BENQO} & \textbf{VQE$^*$} & \textbf{QAOA} \\ 
\hline
\# qubits & $n + 2$ & $n$ & $n$ \\
\# \textsc{cnot} gates & $6m-2n=3n^2+2n$ & $n-1$ & $p\:(2n-2m)=p\:(n^2-n)$ \\
\# single qubit rotations & $2m+n=n^2+2n$ & $n$ & $p\:(m+n)=p\:(n^2+3n)/2$ \\
\# Hadamard gates & $2$ & $2n-2$ & $n$ \\ 
\hline
\# measurement bases & $1$ & $m=n(n+1)/2$ & $m=n(n+1)/2$ \\ 
qubit connectivity & one-to-all & 1 dim.\ nearest neighbor & graph-dependent \\
\hline
\end{tabular}
\end{table*}

Note, that the effective quantum computational resources needed to solve the TSP can be derived by replacing $n$ with ${(n-1)^2}$ in Table \ref{tab:complexity}. This corresponds to a quadratic increase in complexity when dealing with permutation problems, like the TSP, as compared to graph partitioning problems, such as the weighted MaxCut.

\section{Experimental Results}
\label{sec:results}
In this section, we extend the benchmark results presented in Ref.~\cite{Meli.11.06.2023} for the weighted MaxCut problem and evaluate the performance of BENQO for the TSP as defined in \cref{eq:TSPformulation}. The fully-connected graphs used in the experiments were generated with \textsc{NetworkX}~\cite{osti_960616}, with edge weights randomly chosen between $[0,10]$ for MaxCut and $[0,100]$ for the TSP. All quantum circuits were implemented in \textsc{Python} and simulated in the noise-free \texttt{BaseSampler} framework of \textsc{Qiskit}~\cite{Qiskit}, which is based on state vector simulation. 

To expand on previous findings, we increase the number of tested instances from 20 to 100 runs per graph size $n$ for each experiment. Moreover, we simulate the exact probability distributions to achieve results independent of shot noise (that can be mitigated by measuring more often). Further, we replace the small angle approximation of the $\arcsin$ from~\cite{Meli.11.06.2023} with its exact value $\langle\mathbf{C}\rangle=K\arcsin\langle\mathbf{U}\rangle$ (cf.~\cref{eq:UtoC}). 

\subsection{Loss Landscapes}
\label{ssec:LossLandscapes}
\begin{figure*}[!t]
\centering
\includegraphics[width=7.1413in]{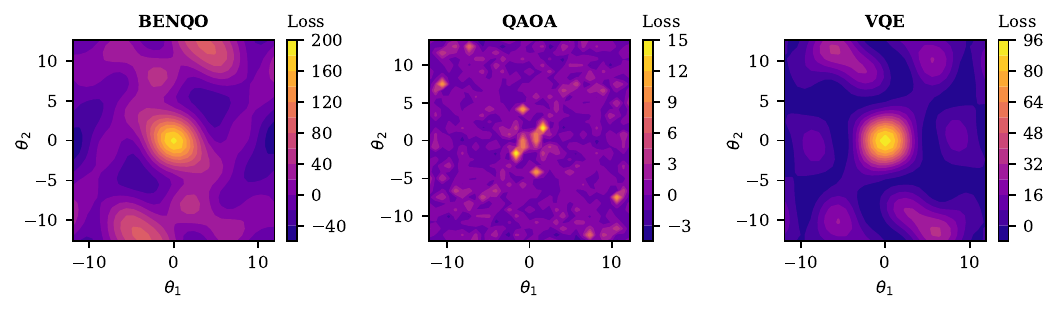}
\caption{Loss Landscapes of three 9-qubit VQAs for the MaxCut Problem with $9$ nodes. The two parameters $\theta_1$ and $\theta_2$ represent the axes of randomly oriented planes in the 10-dimensional (QAOA with 5 layers) and 9-dimensional (BENQO, VQE) parameter space of the tested algorithms.}
\label{fig:losslandscape_MC}
\end{figure*}

\begin{figure*}[!t]
\centering
\includegraphics[width=7.1413in]{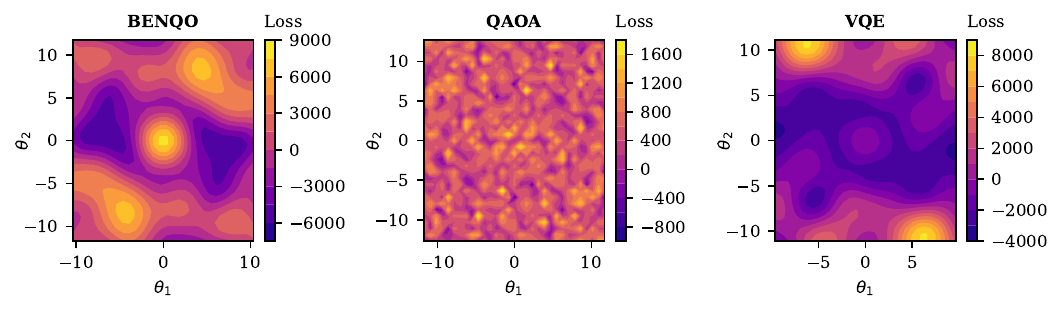}
\caption{Loss Landscapes of three 9-qubit VQAs for the TSP with $3$ nodes. The two parameters $\theta_1$ and $\theta_2$ represent the axes of randomly oriented planes in the 10-dimensional (QAOA with 5 layers) and 9-dimensional (BENQO, VQE) parameter space of the tested algorithms.}
\label{fig:losslandscape_TSP}
\end{figure*}

To intuitively understand how BENQO expresses the cost function within the unitary operator \eqref{eq:unitaryoperator}, \Cref{fig:losslandscape_MC,fig:losslandscape_TSP} illustrate the loss landscapes of BENQO's ansatz with QAOA and a hardware-efficient VQE. All landscapes are based on 9-qubit circuits with 9 (BENQO, VQE) or 10 parameters (QAOA with $p=5$). The cost function is projected onto a randomly oriented plane within the $n$-dimensional hyperspace of the ansatz parameters, bounded within $[-2\pi,2\pi]^n$. The plot variables $\theta_1$ and $\theta_2$ are thus no variational parameters but correspond with a basis of that plane.

QAOA's landscape is filled with closely spaced local extrema and a much smaller range between its maximal and minimal values. Such an energy landscape is highly unfavorable for finding any global solution, especially with gradient-based optimizers. This strongly hints that QAOA's effectiveness heavily relies on the classical optimization strategy and needs good initial points~\cite{Lubinski.05.02.2023}. 

Both BENQO and VQE exhibit a coarser structure with large and meaningful gradients. In addition, the variation of the loss landscape for BENQO and VQE happens at approximately the same scale. This is not surprising, as the individual parameters both control single-qubit rotation gates (unlike QAOA). Therefore, VQE can efficiently search the solution space, but does so without any problem knowledge inherent to the quantum circuit. BENQO, on the other side, uses a problem-specific ansatz, similar to QAOA, via the block encoding of the cost function. It may therefore merge the strengths of both approaches.

Lastly, a brief comment on initialization strategies: for both MaxCut and the TSP, there exists a pronounced maximum at the center of BENQO's and VQE's landscape $(\theta_1,\theta_2)=(0,0)$. For MaxCut, this is intuitive -- aligning all nodes within a single partition yields the maximum cost, as indicated by \eqref{eq:IsingCostExpectation}. As for the TSP, initializing all parameters to zero disrupts all permutational constraints, incurring significant penalties and thus increasing the energy. Consequently, adopting a zero-initialization approach, as practiced in~\cite{Meli.11.06.2023}, appears suboptimal\footnote{In fact, using the exact probability distribution would have immediately ended the optimization process of a gradient-based optimizer because the gradient vanishes due to symmetry. Meli et al.~\cite{Meli.11.06.2023} circumvented this by using a finite number of shots and an approximate value for $\langle\mathbf{C}\rangle$, introducing gradient inaccuracies that prevented such an issue.}. We chose a different strategy for BENQO by drawing the initial points from a normal distribution centered around this critical point.

\subsection{Performance Benchmarking}
In order to assess the optimization performance and effectiveness of BENQO, we benchmark against QAOA and VQE. For a better comparison, each ansatz is chosen with a comparable number of parameters, i.e. $p=\lceil n/2 \rceil$ layers for the QAOA and a simple single-layer two local ansatz (explained in Section \ref{ssec:ressourceandcomplexity}) for the VQE. For the benchmark algorithms, the default versions implemented in \textsc{Qiskit}, including their default initialization strategies, are used. 

Within the BENQO framework, the normalized gradient descent (NGD) technique, detailed in Section \ref{ssec:ParameterOptimization}, is adopted for parameter optimization, limited to a maximum of 20 steps as recommended in Ref.~\cite{Meli.11.06.2023}. This approach was applied not only to BENQO, but also evaluated for QAOA and VQE to assess their efficacy against other optimization strategies. For QAOA, we also explored the use of the well-established COBYLA optimizer~\cite{Powell.1998}, and for VQE, we chose the Powell optimizer~\cite{Powell.1964}. This approach allowed us to compare the impact of gradient-based and gradient-free optimization methods across the examined algorithms. 

To empirically evaluate the performance of all considered methods, a suite of well-established benchmark metrics is employed. The most prominent one for approximation tasks is the approximation ratio
\begin{equation}
    AR = \frac{\langle\psi^*|C|\psi^*\rangle-C_{\text{max}}}{C_{\text{min}}-C_{\text{max}}} \ \in\ [0,1].
\label{eq:approximationratio}
\end{equation}
Here, $|\psi^*\rangle$ is the approximate solution state obtained through the hybrid algorithm. This metric expresses how far the discovered solution differs from the optimal state $|q^*\rangle$ in terms of the energy of the underlying quantum system. The minimum energy and thus cost function is given by $C_{\text{min}}=\langle q^*|C|q^*\rangle$. All terms in \eqref{eq:approximationratio}, including the expected value of the resulting cost, can be determined classically since only diagonal Hamiltonians are involved.

In line with~\cite{Meli.11.06.2023}, the approximation index is a different metric that measures how good the solution represented by the most probable bit string is. It indicates whether the cost of the most probable basis state $|\psi_{\text{max}}\rangle$ in the solution lies within a given threshold $t \in [0,1]$ of the optimum $C_{\text{min}}$:
\begin{equation}
    AI_t = 
    \begin{cases}
    1, & \text{if } \frac{\lvert\langle\psi_{\text{max}}|C|\psi_{\text{max}}\rangle-C_{\text{min}}\rvert}{\lvert C_{\text{min}} \rvert} \leq t \\
    0, & \text{otherwise}
\end{cases}
\label{eq:approximationindex}
\end{equation}
For $t=0$, $AI_0 = 1$ if and only if the most probable basis state $|\psi_{\text{max}}\rangle$ is the optimal solution. In the following, the percentage of optimal solutions is referred to as "optimality". Additionally, we monitor the frequency of solutions within a $5\%$ and $1\%$ margin of the optimal ($t=0.05$ and $t=0.01$ respectively), labeling this measure as "success rates" as suggested by~\cite{Khumalo.2021}. 

\subsubsection{Weighted Maximum Cut Problem}
Expanding on the promising results of the original BENQO proposal~\cite{Meli.11.06.2023} in comparison with QAOA and quantum annealing, we include a second gate-based solving paradigm instead of annealing, namely a simple hardware-efficient VQE ansatz. In the subsequent section, the following solution paths are evaluated: 

\begin{itemize}
    \item BENQO with the NGD optimizer
    \item QAOA (of $p=\lceil n/2 \rceil$ layers) \\with NGD and COBYLA optimizer
    \item VQE (with one-layered two-local ansatz) \\with NGD and Powell optimizer
\end{itemize}

\begin{figure*}[!t]
\centering
\includegraphics[width=7.1413in]{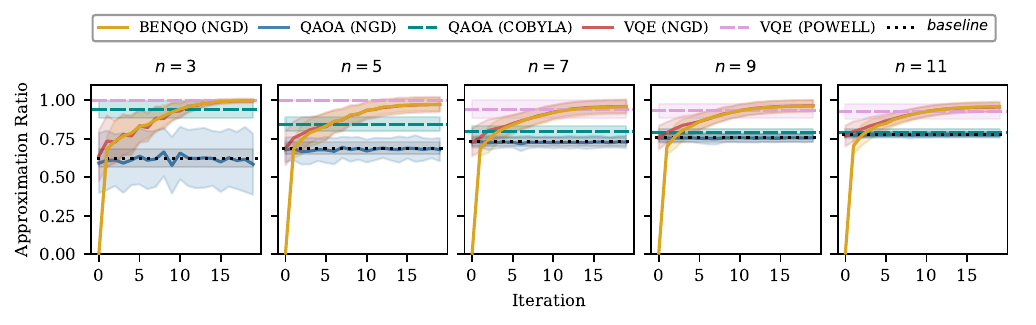}
\caption{Evolution of the approximation ratio across optimization iterations for the MaxCut problem on fully-connected graphs with $n=3,5,7,9$ and $11$ nodes. Dashed lines indicate average ratios for 100 instances using gradient-free methods, i.e. QAOA with COBYLA, and VQE with POWELL. The dotted baselines represent the averaged ratios for the uniform superposition state solution.}
\label{fig:bm_maxcut_ratios}
\end{figure*}

Figure \ref{fig:bm_maxcut_ratios} shows the averaged approximation ratios of 100 optimization runs for fully-connected MaxCut instances with $n=3, 5, 7, 9,$ and $11$ nodes. As a baseline, the mean ratios for the uniform superposition state are additionally included. Our results confirm BENQO's superiority over both QAOA variants while maintaining robust performance even as the number of nodes increases -- in contrast to the fluctuating performance of the QAOA solvers across different system sizes. 
When paired with the gradient-based NGD solver, QAOA performs especially poorly, matching the baseline performance of the uniform superposition state. This aligns with the conjecture made in Section \ref{ssec:Algorithm Outline}.
VQE exhibits a performance close to that of BENQO for both optimizers. Interestingly, VQE shows a potentially better initialization, as indicated by its non-zero start in approximation ratios. This can be attributed to BENQO being initialized near a local maximum, suggesting that adjustments of this initial point strategy may enhance performance. Additionally, the zigzag pattern observed with the path of VQE (NGD) versus the smoother progression of BENQO suggests that VQE may be less compatible with the particular gradient-based optimization method proposed in Ref.~\cite{Suzuki.}, a hypothesis supported by our analysis of loss landscapes in Section \ref{ssec:LossLandscapes}. Adjusting the step size could likely make VQE's path smoother and improve compatibility with this optimization approach.

\begin{figure}[!t]
\centering
\includegraphics[width=1\linewidth]{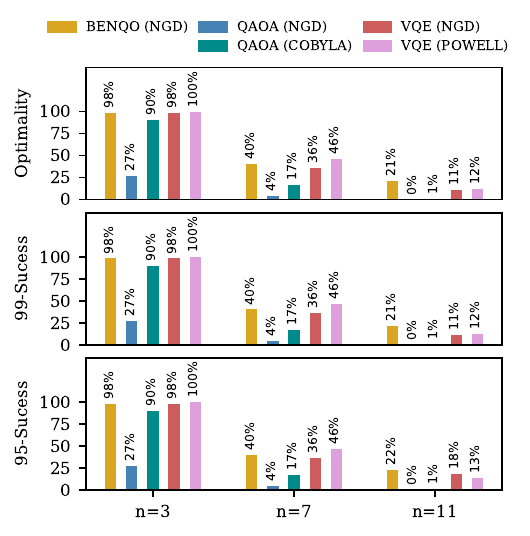}
\caption{Comparison of the Optimality (top) and Success Rates of $99\%$ (middle) and $95\%$ (bottom) of the solution, indicating the percentage of runs in which a globally optimal or close-to-optimal solution was found. Each bar represents the cumulated approximation index of 100 runs for the MaxCut problem with fully-connected graphs of $n=3,7$ and $11$ nodes.}
\label{fig:bm_maxcut_indices}
\end{figure}

Moving on to the optimality and success rates of the 100 test runs, which are shown in Figure \ref{fig:bm_maxcut_indices}, a recurring theme is observable: QAOA, particularly with the NGD solver, has difficulties in achieving (near) optimal solutions, whereas BENQO and VQE demonstrate superior performance. Notably, for $n=11$ nodes, BENQO even achieves slightly higher success rates than both VQE variants. However, unlike the observations made regarding approximation ratios, the performance in terms of optimality does not remain steady as the number of nodes increases. Instead, the proportion of globally optimal solutions declines with larger system sizes, a trend also documented in the original reference~\cite{Meli.11.06.2023}. This is intuitive as high-quality solutions occupy an increasingly smaller fraction of the state space with a larger system size.

\vfill\null
%\vspace{10pt}
\subsubsection{Traveling Salesperson Problem}
To address a more complex and practically relevant challenge, our benchmarking study is extended to the TSP. The chosen encoding (see Section \ref{ssec:TravelingSalesmanProblem}) results in a scenario where not all potential solutions are also feasible. Infeasible solutions are useless for practical purposes and cannot be easily interpreted as approximations of feasible solutions. Many practically relevant optimization problems suffer from this difficulty. To quantify the overlap between the proposed solution $|\psi^*\rangle$ and the set of feasible solutions, we use the feasibility ratio
\begin{equation}
    FR = \sum_{f \in F}|\langle f |\psi^*\rangle|^2
\label{eq:feasibilityratio}
\end{equation}
where $F$ denotes the set of all feasible solutions. In practical settings, where a finite number of measurement shots are used instead of an exact probability simulation, $FR$ can simply be computed as the proportion of feasible shots among total shots.

\begin{figure*}[!t]
\centering
\includegraphics[width=7.1413in]{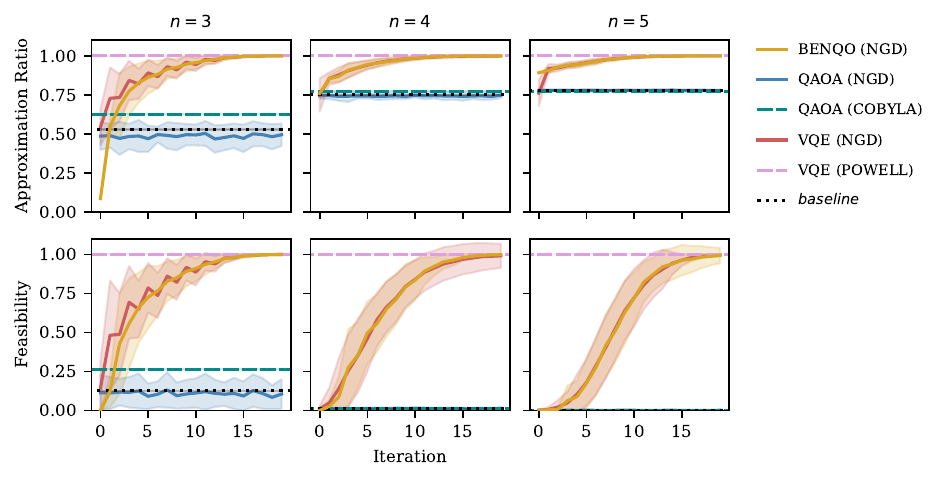}
\caption{Evolution of the approximation ratio (top) and feasibility ratio (bottom) across optimization iterations for the TSP on fully-connected graphs with $n=3,4$ and $5$ nodes. Dashed lines indicate average ratios for 100 instances using gradient-free methods, i.e. QAOA with COBYLA, and VQE with POWELL. The dotted baselines represent the averaged ratios for the uniform superposition state solution.}
\label{fig:bm_TSP_ratios}
\end{figure*}

Figure \ref{fig:bm_TSP_ratios} compares the development of approximation and feasibility ratios throughout the optimization processes of the five methods being analyzed. Each path is averaged across 100 fully-connected graph instances of $n=3,4$ and $5$ nodes. For the TSP, this correlates to circuit complexities of $4$-, $9$-, and $16$-qubits due to the effective problem size of $(n-1)^2$, as detailed in Section \ref{ssec:TravelingSalesmanProblem}). Mirroring trends observed in the MaxCut analysis, BENQO consistently outperforms both QAOA variants and can compete with the two hardware-efficient VQE approaches. Again, the solutions of QAOA (NGD) are the worst, matching the baseline of the uniform superposition state. Also, VQE's characteristic reoccurring zigzag pattern when using normalized gradient descent (NGD) reaffirms the preliminary assumption that gradient-based optimization may not be ideally suited for VQE. 

Furthermore, Figure \ref{fig:bm_TSP_ratios} shows that BENQO and VQE consistently converge towards feasible solutions even as the number of TSP nodes increases. In contrast, QAOA faces challenges from the start -- a trend also observed by~\cite{Khumalo.2021} and~\cite{Palackal.19.04.2023}. With penalty terms incorporated into the energy function \eqref{eq:TSPformulation}, the loss landscape of QAOA is characterized by numerous local minima and barren plateaus, complicating the achievement of feasible solutions within the usual parameter range \cite{Xie.17.08.2023}. Palackal et al.~\cite{Palackal.19.04.2023} conjecture that reaching a specific energy threshold is essential for a solver to find feasible solutions. According to the loss range displayed in Figure+\ref{fig:losslandscape_TSP}, this target proves challenging for the QAOA. As a result, it often yields solutions with low feasibility for constrained problems.
This situation presents an opportunity for BENQO, which directly encodes the problem, unlike VQE, which in its ansatz almost resembles a random state generator. Despite this difference, it consistently yields feasible solutions and demonstrates convergence properties comparable with those of VQE, suggesting a potential advantage in its application.

\begin{figure}[!t]
\centering
\includegraphics[width=1\linewidth]{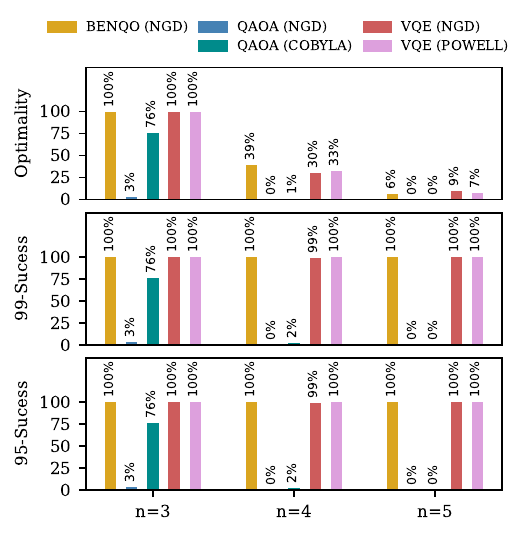}
\caption{Comparison of the Optimality (top) and Success Rates of $99\%$ (middle) and $95\%$ (bottom) of the solution, indicating the percentage of runs in which a globally optimal or close-to-optimal solution was found. Each bar represents the cumulated approximation index of 100 runs for the TSP with fully-connected graphs of $n=3,4$ and $5$ nodes.}
\label{fig:bm_TSP_indices}
\end{figure}

The optimality and success rates in \Cref{fig:bm_TSP_indices} support these findings. QAOA faces similar challenges in reaching optimal solutions for the TSP as it does with MaxCut, whereas BENQO matches or occasionally surpasses the high optimality rates of VQE. For these two algorithms, success rates largely remain at or near $100\%$ despite the quadratic increase in the number of qubits for the TSP. This contrasts with the MaxCut analysis (see Figure \ref{fig:bm_maxcut_indices}), where success rates significantly declined with increasing problem size. This change in behavior can be attributed to the scaling of the cost function with penalty terms in \eqref{eq:TSPformulation}, increasing the overall cost range. As a result, a larger proportion of solutions qualifies as optimal or nearly optimal for the TSP when examining energies or, equivalently, approximation ratios. This underlines the importance of a problem-specific view on algorithm performance.

Finally, to provide another measure that reflects application-specific performance, the following length ratio based on Ref.~\cite{Palackal.19.04.2023} is adopted:
\begin{equation}
LR = \frac{l_{\text{TSP}}-l_{\text{max}}}{l_{\text{opt}}-l_{\text{max}}}\ \in \ [0,1].
\end{equation}
Here, $l_{\text{TSP}}$ is the weighted average path length over all feasible states of the solution proposal, and the optimal path length $l_{\text{opt}}$ is determined by the shortest path found via the best-known classical methods or through brute-force calculation for smaller instances. This ratio offers another quantifiable measure of how closely the algorithm approximates the optimal solution within the context of the TSP. However, it discards all infeasible states and thus has to be interpreted in conjunction with the feasibility ratio.

Table \ref{tab:length_ratios_tsp} presents these length ratios for TSP instances with $n=4$ and $5$ nodes solved by the five considered methods -- for $n=3$, the measure is trivially $1$ since there is only one cycle to begin with. Additionally, a baseline for the average $LR$ produced by the uniform superposition state over $(n-1)^2$ qubits was added. As the length ratio is a more reliable measure for solution optimality, one can again observe a declining trend as the number of nodes increases, which is consistent with preceding MaxCut results. Once more, BENQO closely competes with VQE, while QAOA lags behind in achieving comparable outcomes. Note that the standard deviations of BENQO's and VQE's results are much higher than those for QAOA and the uniform superposition state, indicating that they reached much higher values for a non-significant proportion of the runs. This pattern reinforces the significance of selecting appropriate optimality measures to accurately assess and compare solver efficiency, especially in the context of increasing problem complexity.

\begin{table}[!t]
\renewcommand{\arraystretch}{1.2}
\caption{Length Ratios averaged over 100 TSP instances with $n=4$ and $5$ nodes, for BENQO, four benchmark quantum optimizers, and a baseline for the uniform superposition state.}
\label{tab:length_ratios_tsp}
\centering
\begin{tabular}{ll|llll}
\multicolumn{1}{c}{} & \multicolumn{1}{c|}{} & \multicolumn{2}{c}{\textbf{Length Ratio}} \\ \hline
\multicolumn{1}{l|}{\textbf{Algorithm}}    & \textbf{Optimizer}    & \multicolumn{1}{c}{$n=4$}  & \multicolumn{1}{c}{$n=5$}  \\ \hline
\multicolumn{1}{l|}{BENQO} & NGD  & $0.561
\pm0.439$ & $0.493\pm0.336$ \\ \hline
\multicolumn{1}{l|}{\multirow{2}{*}{QAOA}} & NGD & $0.554\pm0.181$ & $0.505\pm0.075$ \\
\multicolumn{1}{l|}{} & COBYLA & $0.554\pm0.203
$ & $0.505\pm0.090$ \\ \hline
\multicolumn{1}{l|}{\multirow{2}{*}{VQE}} & NGD & $0.579\pm0.431$ & $0.466\pm0.330$ \\
\multicolumn{1}{l|}{} & POWELL  & $0.601\pm0.445
$ & $0.525\pm0.322$ \\ \hline
\multicolumn{2}{l|}{*uniform superposition state*} & $0.562\pm 0.088$ & $0.500\pm 0.000$
\end{tabular}
\end{table}

\subsection{Runtime Analysis}
Finally, an analysis of the total time to solution of the different solver instances is conducted to gain insights into BENQO's efficiency. As suggested in~\cite{Khumalo.2021}, the average CPU time without outliers is used for this study. Outliers are employed via the $z$ score
\begin{equation}
    z(x_i) = \frac{x_i-\boldsymbol{\overline{x}}}{\Delta\boldsymbol{x}}
\label{eq:zscore}
\end{equation}
and removed if $z(x_i) > 1.5$. Exact timing results would vary significantly across the diverse high-performance computing systems utilized in research. In the following, we therefore aim for a qualitative and comparative analysis to gain an initial understanding.

\begin{figure}[!t]
\centering
\includegraphics[width=1\linewidth]{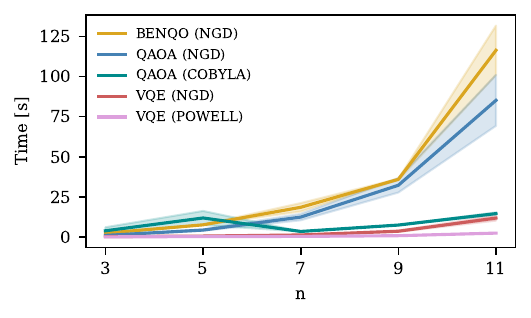}
\caption{CPU Runtime (to solution) averaged over approx. 100 runs without outliers for MaxCut instances of $n=3,5,7,9$ and $11$ nodes.}
\label{fig:bm_maxcut_time}
\end{figure}

\begin{figure}[!t]
\centering
\includegraphics[width=1\linewidth]{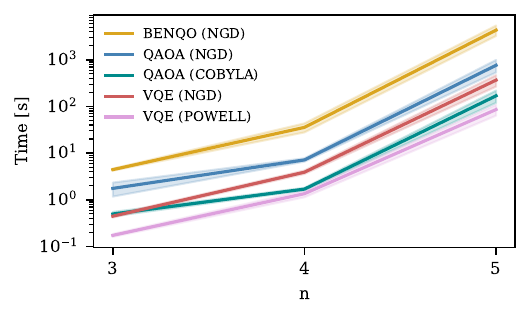}
\caption{CPU Runtime (to solution) averaged over approx. 100 runs without outliers for TSP instances of $n=3,4$ and $5$ nodes. Times are plotted on a logarithmic scale to account for the quadratic increase in problem size.}
\label{fig:bm_tsp_time}
\end{figure}

\Cref{fig:bm_maxcut_time,fig:bm_tsp_time} show the average runtimes for the MaxCut and TSP experiments as the number of nodes increases. BENQO emerges as the slowest solver compared to the other benchmarked optimizers, with VQE (POWELL) being notably the fastest. Despite these speed differences, all solvers demonstrate a similar growth pattern in runtime as problem sizes increase. No solver struggles disproportionately with larger problem sizes. Interestingly, all algorithms using the gradient-based optimizer perform more slowly compared to their gradient-free counterparts, despite the latter needing more iterations. This can be attributed to the higher number of circuit executions required to calculate gradients, i.e., $2$ per parameter using the parameter shift rule \eqref{eq:parametershiftrule}. 

Several factors may contribute to BENQO's consistently slower performance:
First, BENQO's block-encoded circuits exhibit a quadratic increase in depth with the problem size $n$, as detailed in Table \ref{tab:complexity}, resulting in circuit depths much greater than those of the linearly scaling, hardware-efficient VQE. At the same time, Meli et al.~\cite{Meli.11.06.2023} show that BENQO holds a gate-count advantage over QAOA when the number of QAOA layers exceeds $3$. This is the case for MaxCut with $n\geq6$ or TSP with $n\geq4$ (or $\geq9$ qubits) in our simulations. Yet, this theoretical advantage does not manifest in our experiments. The deviation probably stems from the fact that unlike the well-established solvers implemented within \textsc{Qiskit}, BENQO has not been fine-tuned for computational efficiency, yet.

\section{Discussion and Conclusion}
\label{sec:discussion}
%% Summary
In this study, we investigated BENQO, a novel hybrid quantum-classical optimization algorithm~\cite{Meli.11.06.2023}. It uses block encoding to implement the cost function into a quantum circuit and the principle of implicit measurement for efficient cost expectation sampling. For the weighted MaxCut and the more complex Traveling Salesperson Problem, BENQO's performance is benchmarked against two well-known quantum algorithms, QAOA and VQE, using both a gradient-based and a gradient-free optimizer. Our evaluation covered various aspects such as computational resource costs, problem complexity, solution quality, and runtime, as suggested in Ref.~\cite{Abbas.04.12.2023}, and a comparative analysis of loss landscapes. 

%% Key Findings
BENQO's superior performance over QAOA is confirmed by problem-agnostic and problem-specific benchmark metrics. BENQO shows improved convergence behavior and a better ability to generate feasible and optimal solutions. 
Furthermore, BENQO stands out as a strong competitor to VQE, achieving comparable results in the key performance metrics. The tested hardware-efficient VQE ansatz, especially with the Powell solver, demonstrated very good performance in all tested categories. However, BENQO may also have significant advantages:

%% Interpretation / Implications / Relevance
While the selected VQE ansatz might be seen as just a sophisticated random state generator, BENQO embeds problem specifics into its ansatz, similar to QAOA. Yet, it exhibits clear and optimizable energy landscapes like VQE, leading to the same high-quality solutions. This suggests that BENQO may unite the strengths of both QAOA and VQE, making it a flexible and powerful algorithmic paradigm, which should be explored further.

%% Limitations / Suggestions for Improvement / Call for Action
BENQO is clearly able to yield high-quality solutions, but in terms of its simulated runtime, it remains behind VQE. However, BENQO's circuits are neither decomposed nor optimized prior to execution, likely resulting in the extended simulation times observed in our experiments. In conclusion, optimizing BENQO's circuits and selecting more effective parameter optimizers present considerable opportunities to improve its efficiency.

%% Final Statement (my contribution)
In conclusion, this study shows the potential of BENQO as a promising candidate for solving combinatorial optimization problems through hybrid quantum-classical computing. By demonstrating its strengths -- even for a complex and practically relevant problem like the TSP -- we hope to inspire further research in this dynamic field. Further developing BENQO may not only unlock its full potential but also pave the way for new advancements in variational quantum optimization beyond QAOA and VQE.

\balance
\section{Outlook}
\label{sec:outlook}
While this study has unveiled substantial insights, offering a promising new solution paradigm tailored for the NISQ era, various aspects of BENQO call for further exploration. As highlighted before~\cite{Abbas.04.12.2023}, in this advancing era, progress in quantum algorithms must come from both theoretical analysis and empirical methods. 

On the empirical front, a comprehensive optimization of BENQO's hyperparameters is therefore suggested, including the fine-tuning of initial parameter settings, penalty factors, classical optimizer choices, and the ansatz structure itself, to enhance BENQO's performance. 
Additionally, evaluating the influence of the sampling rate on BENQO's results and more thoroughly investigating the potential benefits of the specific measurement technique rooted in the principle of implicit measurement would be interesting avenues for further research.
Our runtime analysis also indicates an urgent need to explore circuit optimization techniques aimed at reducing both the depth and runtime of BENQO circuits. Currently, the potential circuit decompositions suggested in Ref.~\cite{Meli.11.06.2023} remain unexplored. Implementing these strategies could significantly accelerate computational processes, thereby enabling the extension of our analysis to tackle more complex problems, such as the TSP with up to 6 nodes.

Additionally, investigations of a theoretical nature are recommended.
This involves looking into possible performance guarantees, better understanding algorithmic complexity, and statistically analyzing loss landscapes as detailed by~\cite{Rajakumar.2024}. These steps are critical for building a solid theoretical base to support BENQO's performance.
Following the advice in Ref.~\cite{Abbas.04.12.2023}, an exploration of problem types beyond typical QUBO and Ising could further yield valuable insights. A possible investigation could involve applying binary encodings to the TSP, moving towards Higher-Order Binary Optimization (HOBO)~\cite{Salehi2022, Dominguez.2023}. Adapting BENQO to HOBO problems would test its effectiveness on a broader range of challenges, which could open up new research directions.



\clearpage
\balance %somewhere on the last page, to balance columns
%\bibliographystyle{IEEEtran}
%\bibliography{reference}
\printbibliography
\end{document}